\documentclass[prB,aps,twocolumn,showpacs]{revtex4-1}

\usepackage{graphicx}
\usepackage{bm,color}
\usepackage{physics}
\usepackage{amsmath}
\usepackage{bm,color}
\usepackage{multirow}

\newcommand{\be}{\begin{eqnarray}}
\newcommand{\ee}{\end{eqnarray}}

\newcommand{\Se}{S_{\rm ent}}

\begin{document}

\title{
Non-thermalized dynamics of flat-band many-body localization}
\date{\today}
\author{Takahiro Orito$^{1,2}$}
\author{Yoshihito Kuno$^{3,4}$}
\author{Ikuo Ichinose$^1$}
\affiliation{$^1$Department of Applied Physics, Nagoya Institute of Technology, Nagoya, 466-8555, Japan}
\affiliation{$^2$Department of Physics, Graduate School of Science, Hiroshima University}
\affiliation{$^3$Department of Physics, Graduate School of Science, Kyoto University, Kyoto 606-8502, Japan}
\affiliation{$^4$Department of Physics, Graduate School of Science, Tsukuba University, Tsukuba, Ibaraki 305-8571, Japan}

\begin{abstract}
We find that a flat-band fermion system with interactions and without disorders exhibits non-thermalized
ergodicity-breaking dynamics, an analog of many-body localization (MBL). 
In the previous works, we observed flat-band many-body localization (FMBL) in the Creutz ladder model.
The origin of FMBL is a compact localized state governed by local integrals of motion (LIOMs), 
which are to be obtained explicitly. 
In this work, we clarify the dynamical aspects of FMBL.
We first study dynamics of two-particles, and find that the states are not substantially modified by
weak interactions, but the periodic time evolution of entanglement entropy emerges 
as a result of a specific mechanism inherent in the system. 
On the other hand, as the strength of the interactions is increased, the modification of the states takes place 
with inducing instability of the LIOMs. 
Furthermore, many-body dynamics of the system at finite fillings is numerically investigated by time-evolving block decimation (TEBD) method. For a suitable choice of the filling, 
non-thermal and low entangled dynamics appears.
This behavior is a typical example of the disorder-free FMBL.
\end{abstract}


\maketitle
\textit{ Introduction.---}
At present, many-body localization (MBL) is one of the most intensively studied topics, both theoretically and experimentally \cite{Nandkishore,Abanin,Imbrie}.
Despite the numerous studies, however, essential properties of MBL and its transition remain an open question.
It is now widely accepted that there emerge an extensive number of local conserved quantities in 
MBL states \cite{Nandkishore}, which are called local integrals of motion (LIOMs) \cite{Serbyn,Huse}, but how the LIOMs 
behave at transitions out of MBL states is not clear \cite{Morningstar,Vasseur}. 
The LIOMs are the number operator of local bits ($\ell$-bits), and in most of the studies on the LIOMs given so far,
$\ell$-bits are constructed, e.g., from single-particles states of Anderson localization. 
Therefore, they are numerically obtained 
\cite{Bera,Imbrie1,Rademarker1,You1,Pekker1,Brien1,Wahl1,Pal1,Tomasi,Peng,Nicolas}
and depend on the random potential originating Anderson localization.
Study on systems with explicit form of $\ell$-bits is certainly desired and welcome.

In the previous works \cite{KOI2020,OKI2020}, we investigated Creutz ladder model \cite{Creutz1999,Bermudez,Junemann,Tovmasyan,Sun,Barbarino_2019} with interactions, 
in which all single particle spectra are flat, for motivation finding MBL-like behaviors.
In fact by using exact diagonalization, this expectation was verified \cite{KOI2020,Roy}. 
General discussion of the presence of flat-band MBL (FMBL) has been reported \cite{Danieli_1}.
The origin of localization in that system comes from {\it compact localized states} \cite{Leykam,Zurita,Zhou,kuno2020}, 
which can be regarded as $\ell$-bits from the view point of MBL.  
In the single-particle level,
the number operator of the compact-localized state is LIOMs. 
All energy eigenstates in the model have a finite support and the energy level is degenerate in the non-interacting limit. 
Even in the presence of interactions, one may expect that the effects of the single-particle LIOMs remain and they
make the system localized \cite{KOI2020}. 
We call this phenomenon FMBL. 
Signature of FMBL has been reported not only for the Creutz ladder but also other flat-band models \cite{Roy,Zurita,Paul,Daumann,Tilleke,Khare}.
However, its dynamical properties have not been clarified so far. 
In particular, dynamical entanglement properties of 
FMBL is not completely understood, though some studies on it have been given, e.g., from
the perspective on the Aharanov-Bohm caging \cite{vidal0,vidal1} with interactions \cite{Liberto,Danieli1,Danieli2,KMH2020}. 
In this Letter, we clarify the dynamics of FMBL for {\it Creutz-ladder fermions with the explicit LIOMs}
by observing the quantities employed 
in studying the conventional MBL states~\cite{Nandkishore,Abanin,Imbrie}. 

First, we study two-particle dynamics to elucidate the entanglement properties of the system
since the $\ell$-bit picture is significantly clear there. 
With interactions, entanglement entropy ($S_{\rm ent}$) is produced, whose behavior gives an insight into 
the many-body dynamics of FMBL. 
We then investigate many-body states at finite fillings. 
We find a FMBL state that exhibits non-thermalizing behavior, i.e., memory of an initial state is preserved, 
and also a nontrivial slow growth of $S_{\rm ent}$ appears, which is not logarithmic growth as in the conventional MBL 
state \cite{Bardarson}. 
The stability of non-thermal dynamics depends on the particle filling.
We find that it is determined by robustness of the  $\ell$-bit picture.

\textit{Model.---}
Target model is the Creutz ladder system with Hamiltonian given as follows,
\be
H_{\rm CL} = && \sum^{L}_{j=1}\Big[-i\tau_1 (a^\dagger_{j+1}a_j-b^\dagger_{j+1}b_j) \nonumber \\
&&\hspace{1cm} -\tau_0(a^\dagger_{j+1}b_j+b^\dagger_{j+1}a_j)+ \mbox{h.c.} \Big],
\label{HCL}
\ee
where $a_j$ and $b_j$ are fermion annihilation operators at site $j$, and $\tau_1$ and $\tau_0$ are
intra-chain and inter-chain hopping amplitudes, respectively. 
The corresponding system has been experimentally realized \cite{Shin2020}.
For $\tau_1=\tau_0$, the system has two bands, which are strictly flat, 
and the perfect localization is realized.
The Hamiltonian (\ref{HCL}) is written as a summation of LIOMs such as 
$H_{\rm flat}\equiv H_{\rm CL}|_{\tau_1=\tau_0}$ with the $\ell$-bits $W^{\pm}_j$, 
\be
H_{\rm flat} &=& \sum^{L}_{j=1} [-2\tau_0W^{+\dagger}_j W^+_j+2\tau_0W^{-\dagger}_jW^-_j],
\label{HWW}  \\
W^+_j&=&{1 \over 2}(-ia_{j+1}+b_{j+1}+a_j-ib_j),  \label{Wab}  \\
W^-_j&=&{1 \over 2}(-ia_{j+1}+b_{j+1}-a_j+ib_j). \nonumber
\ee
It is readily verified
$
\{W^{\alpha \dagger}_j ,W^{\beta}_k \}=\delta_{\alpha, \beta}\delta_{jk}, 
\;\; (\alpha, \beta = \pm), \; \mbox{etc.}
$
In what follows, we set $\tau_{0}=\tau_{1}$. 
Then, all single-particle eigenstates are given as 
$(W^\pm_j)^\dagger |0\rangle$ with energy $\mp 2 \tau_0$. 
A single localized particle exhibits the Aharanov-Bohm caging \cite{vidal0,vidal1} around {\it the plaquette} $j$. 
The particle does not spread out through the system and oscillates around the initial site. 
As a result, exact localized dynamics appears in a single particle level.

From $H_{\rm flat}$ in Eq.~(\ref{HWW}), operators $K^\pm_j \equiv W^{\pm \dagger}_j W^\pm_j$
are nothing but LIOMs, which play an important role in study on localization.
Besides $H_{\rm flat}$ in (\ref{HWW}), the system Hamiltonian can have an inter-leg hopping term. 
The effect of it is studied in \cite{Supp} in order to investigate properties of extended states.

\begin{figure}[t]
\begin{center} 
\includegraphics[width=8cm]{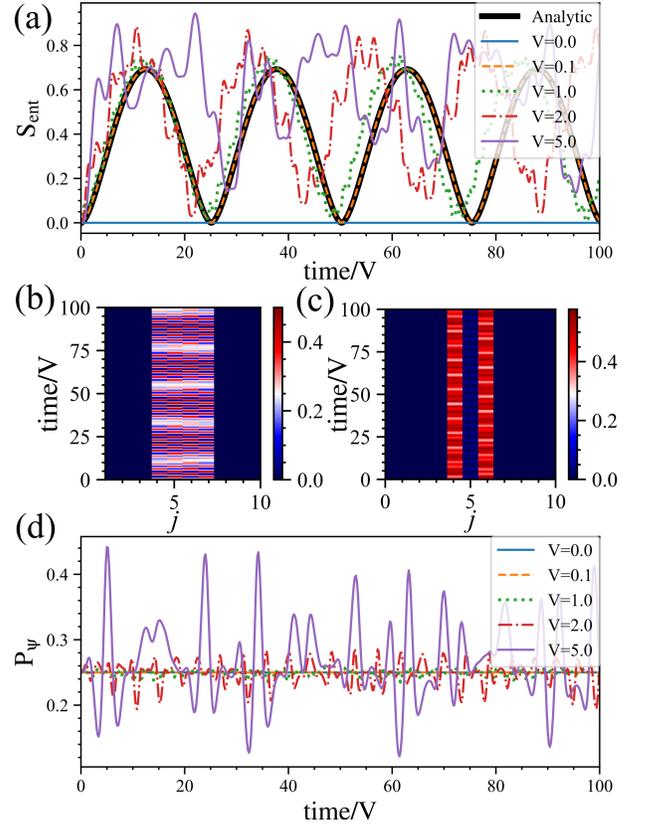} 
\end{center} 
\caption{
Two-particle dynamics starting with initial state;
${1\over 2}(a^\dagger_4+ib^\dagger_4) (a^\dagger_6+ib^\dagger_6)|0\rangle$.
(a) Time evolution of $\Se$, for various $V$'s.
$S_{\rm ent}$ exhibits an oscillating behavior for $V\le 1$, whereas for larger $V$'s, 
its behavior becomes complicated.
(b) Time evolution of $a$-particle density for $V=2.0$.
(c) Time evolution of $\langle K^-_j\rangle$ for $V=2.0$.
Its instability is observed.
(d) Time evolution of the fourth-order moment of LIOMs $P_{\psi}$ for the flat band with various $V$'s.
For larger $V$'s, it oscillates indicating instability of LIOMs, but the revival to the initial value
is also observed. 
For all data, the system size is $L=10$ and we set open boundary condition.
}
\label{Fig1}
\end{figure}

\textit{Two-particle system.---}
Let us first consider two-particle system to investigate effects of the interactions such as 
$H_{\rm I}=V\sum_j (n^a_jn^b_j+n^a_jn^a_{j+1}+n^b_jn^b_{j+1})$, 
where $V>0$ and $n^{a(b)}_j=a^\dagger_ja_j (b^\dagger_jb_j)$.
Study on the two-particle system gives an insight on dynamics of many-body states that we
consider later on.
The target system is given by $H_{\rm flat}+H_{\rm I}$ with the unit of energy 
$\tau_0=1$.
We numerically study $S_{\rm ent}$. The entanglement entropy $S_{\rm ent}$ is defined as the von-Neumann entanglement entropy for a reduced density matrix for a subsystem, 
$S_{\rm ent}=-{\rm Tr}\rho_A \ln \rho_A$, 
where $\rho_A={\rm Tr}_{B}|\Psi\rangle \langle \Psi|$ is a reduced density matrix, $|\Psi\rangle$ is a many-body eigenstate and system is divided into A and B subsystem. 
We also calculate expectation values of particle density and LIOMs, 
$\langle K^\pm_j \rangle$ during time evolution. 
As well as the above quantities, we introduce the fourth-order moment of LIOMs, $P_\psi(t)=\sum_\pm P^\pm_\psi(t)$, 
where $P^{\pm}_\psi \equiv \sum^{L}_{j=1}|\langle \psi|K^{\pm}_j|\psi\rangle|^4$. 
The fourth-order moment of LIOMs quantifies the stability of the LIOMs in states \cite{remark1}.
In later investigation, its measure indicates {\it local enhancement of $\ell$-bit}, which is an interesting phenomenon.

As in Ref.~\cite{Serbyn2012_2}, we first consider a two-particle system in which each particle state consists 
of a superposition of two states such as 
${1 \over 2}(W^{+\dagger}_j-W^{-\dagger}_j)(W^{+\dagger}_k-W^{-\dagger}_k)|0\rangle = {1\over 2}(a^\dagger_j+ib^\dagger_j) (a^\dagger_k+ib^\dagger_k)|0\rangle$.
This consideration of the two-particle states gives very important insight on FMBL.
In the practical calculation, we put the distance between two states $|j-k|=2$.
For $|j-k|\ge 3$, effects of interactions do not work.
Density matrix and $S_{\rm ent}$ of the $j$-subsystem are defined as usual by taking trace of the $k$-subsystem \cite{remark2}.

In Fig.~\ref{Fig1} (a), we show observation of time evolution of $S_{\rm ent}$  for various values of $V$ \cite{Quspin}.
Initial state is ${1 \over 2}(W^{+\dagger}_4-W^{-\dagger}_4)(W^{+\dagger}_6-W^{-\dagger}_6)|0\rangle = {1\over 2}(a^\dagger_4+ib^\dagger_4) (a^\dagger_6+ib^\dagger_6)|0\rangle$,
and $\Se$ is obtained by integrating out quantum degrees of freedom on sites $j\ge 6$. 
The initial two-particles start to spread out by Aharanov-Bohm caging and then, the particles interact 
with each other via the nearest-neighbor components of $H_{\rm I}$.

The calculations of $\Se$ exhibit very interesting behavior.
For small $V\le 1$, $\Se$ oscillates periodically, and the period is almost the same for $V$'s smaller than unity. 
We also show the analytical solution of the dynamics of $S_{\rm ent}$ (See \cite{Supp}). 
The numerical results for small $V$ are in good agreement with the analytical result.
One may think that this behavior comes from a similar effect to that in the well-separated two-particle
system studied in Ref.~\cite{Serbyn2012_2}.
However, this is not the case.
By assuming that the states are not modified by the weak interactions, let us focus on the four states,
$
|\alpha\rangle\equiv W^{+\dagger}_4W^{+\dagger}_6|0\rangle, 
|\beta\rangle\equiv W^{+\dagger}_4W^{-\dagger}_6|0\rangle, 
|\gamma\rangle\equiv W^{-\dagger}_4W^{+\dagger}_6|0\rangle, 
$
and 
$
|\delta\rangle\equiv W^{-\dagger}_4W^{-\dagger}_6|0\rangle.
$
In the usual case~\cite{Serbyn2012_2}, the interaction energy between particles located at sites $5$ and $6$
in the $(a,b)${\it -particle representation}
gerenates $S_{\rm ent}$, which exhibits an oscillating behavior in time.
In the present case however, the states $|\beta\rangle$ and $|\gamma\rangle$ are strictly degenerate.
Furthermore, the interaction $H_{\rm I}$ {\it cannot} be expressed solely in terms of the LIOMs, $K^\pm_j$'s,
and there exist terms such as $(W^{+}_jW^{-\dagger}_jW^{+\dagger}_kW^{-}_k+\mbox{h.c.})$, etc.
In fact, it is easily verified $\langle \beta|H_{\rm I}|\gamma\rangle \neq 0$.
Therefore, the mixing between $|\beta\rangle$ and $|\gamma\rangle$ must be suitably taken into
account by considering linear combinations $|\beta\rangle\pm |\gamma\rangle$
to study the time evolution of the states, although the above non-LIOM terms can be
neglected in certain cases \cite{Tomasi,Nicolas}.
\textit{The above properties are very characteristic ones and essential for the time evolution in the flat-band system.}
We have also verified that other mixing between states such as $|\alpha\rangle$ and $|\beta\rangle$
does not plays a substantial role in the time evolution of $S_{\rm ent}$. 
For $V\le 1$, consideration within the space 
$\{|\alpha\rangle, |\beta\rangle, |\gamma\rangle, |\delta\rangle \}$ is legitimate
as modification of states is negligible, and as a result, 
the values of the peaks of $S_{\rm ent}$ in Fig.~\ref{Fig1} (a) is very close to the theoretical value, 
$\ln 2=0.693\cdots$.
Our estimated period of $\Se$ oscillation is $8\pi$, which is also very close to the results in Fig.~\ref{Fig1} (a).

On the other hand for $V=2.0$ and $5.0$, $\Se$ displays rather complicated behavior and it becomes larger
than $\ln 2$ in certain temporal periods.
This indicates that a modification of eigenstates by the interactions takes place,
and states other than $\{|\alpha\rangle, |\beta\rangle, |\gamma\rangle, |\delta\rangle \}$ emerge.
In order to verify the above observation, we display the time evolution of the particle density
and $\langle K^{\pm}_j \rangle$ in Figs.~\ref{Fig1} (b) and (c) for $V=2.0$.
As expected, the particle densities exhibit rather strong oscillating behavior.
Similar behavior is observed for $\langle K^{\pm}_j \rangle$.
Furthermore, the calculation of the fourth-order moment of LIOMs in Fig.~\ref{Fig1} (d) is quite instructive.
In the initial state, 
$\langle K^+_j\rangle=\langle K^-_j\rangle=1/2$ for
$j=4, 6$ and otherwise zero at $t=0$.
Therefore $P_\psi(t=0)=(1/2)^4\times 2\times 2=1/4$.
The calculation of the the fourth-order moment of LIOMs indicates that the initial particle state is stable for $V \le 1$ but
is strongly modified by the interactions for $V>1$ under the time evolution.
The oscillation of $P_\psi(t)$ and $P_\psi(t)>1/4$ at peaks mean that not only destruction of $\ell$-bit
but also certain local enhancement of $\ell$-bit takes place under the time evolution.
At present, if this phenomenon is peculiar properties of the present model is not known.
This is an interesting future problem.
Anyway, modification of states is observed, as the interactions dominate the energy gap between 
the two flat bands for $V\ge 2$.
This phenomenon has not been observed in the past. 

\begin{figure}[t]
\begin{center} 
\includegraphics[width=8cm]{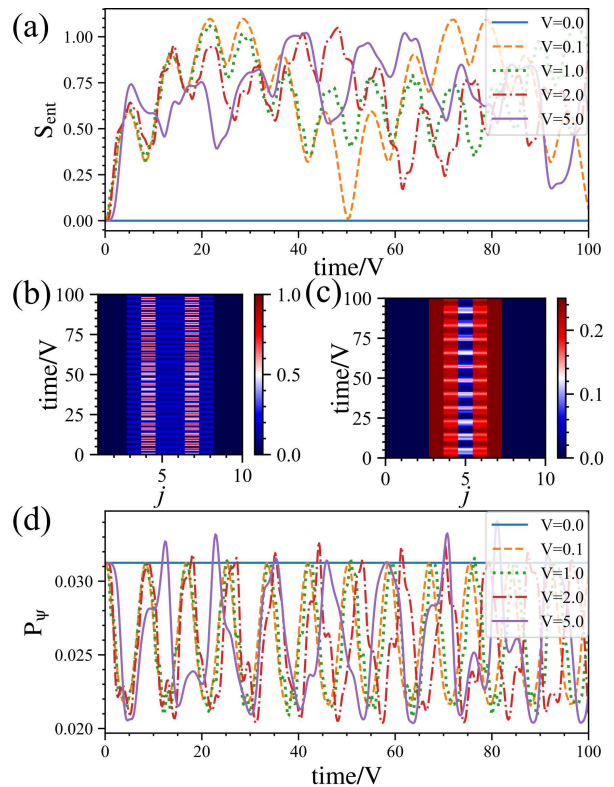} 
\end{center} 
\caption{
Two-particle dynamics starting with initial state; $a^\dagger_4 a^{\dagger}_7|0\rangle$.
(a) Time evolution of $S_{\rm ent}$ for various $V$'s.  
(b) Time evolution of $a$-particle density for $V=2.0$.
(c) Time evolution of $\langle K^-_j\rangle$ for $V=2.0$.
Oscillating behavior with a definite period is observed.
(d) Time evolution of the the fourth-order moment of LIOMs $P_{\psi}$ for various $V$'s. 
For finite $V$, $P_{\psi}$ oscillates strongly indicating instability of LIOMs.
However, the revival to the initial value is observed.
It is interesting to note that locations of maxima (minima) of $S_{\rm ent}$ correspond to minima (maxima) of $P_{\psi}$. For all data, the system size is $L=10$ and we set open boundary condition.}
\label{Fig2}
\end{figure}


As the second case, we prepare an initial state such as, $a^\dagger_4 a^{\dagger}_7|0\rangle$, 
where the distance of the initial two-particles is three site, $|j-k|=3$. 
This configuration is closely related with a many-body state studied later on.
The initial two-particles again spread first by Aharanov-Bohm caging and then, the particles interact with each other by $H_{\rm I}$. 
$S_{\rm ent}$ of the two-particles for various $V$'s is plotted in Fig.~\ref{Fig2} (a) \cite{remark2}. 
While the particles are not entangled in the absence of interactions, 
$S_{\rm ent}$ exhibits clear oscillation for finite $V$ and no saturation. 
For $V\le 1.0$, $S_{\rm ent}$ exhibits an oscillation with multiple periods.
This behavior is different from that of the first case, but it stems from the fact that the initial state
 $a^\dagger_4 a^{\dagger}_7|0\rangle$ is a superposition of $4\times 4=16$ eigenstates of the flat
Hamiltonian, $H_{\rm flat}$, 
$\{W^{+\dagger}_3, \cdots, W^{-\dagger}_4\}\times \{W^{+\dagger}_6, \cdots, W^{-\dagger}_7\}|0\rangle$ (See \cite{Supp}).
Therefore, various resonant states emerge in the time evolution similar to the first case.
On the other hand for $V=2.0$ and $5.0$, $S_{\rm ent}$ shows rather complicated behavior.
This indicates a modification of eigenstates by the interactions.
However, the values of the peaks of $S_{\rm ent}$ are in between $\ln 2$ and $1.1 < \ln 4$.

As the density profile in Fig.~\ref{Fig2} (b) shows, two-particles do not spread out
from the sites $j=3 \sim 8$ for $V=2.0$. 
The LIOMs distribution $\langle K^{-}_j\rangle$ in Fig.~\ref{Fig2} (c) for $V=2.0$ has finite value in dynamics, 
while it oscillates nontrivially. 
Thus, the $\ell$-bit picture survives even for the $V=2.0$ interactions.
We have verified a similar behavior for $V=5.0$.

The calculation of the fourth-order moment of LIOMs in Fig.~\ref{Fig2} (d) shows how the LIOM picture changes in the time evolution. 
The fourth-order moment of LIOMs for finite $V$'s fluctuates but revivals to the initial value, $(1/4)^4 \times 4 \times 2=1/32=0.03125$. 
This indicates that the initial LIOM picture governs the dynamics and keeps the particles localized
for a long period.
For a homogeneous state at finite fillings,
it is plausible to expect that the repulsions between particles suppresses the density fluctuations
and enhances localization,
and then the $\ell$-bit picture and LIOMs work effectively to describe localized states.
We shall verify this expectation in the following.


\begin{figure}[t]
\begin{center} 
\includegraphics[width=7.5cm]{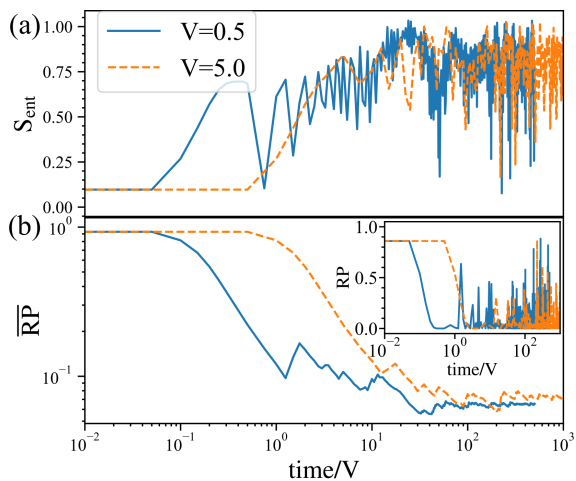} 
\end{center} 
\caption{
$1/6$-filling dynamics: (a) Time evolution of $S_{\rm ent}$ for subsystem with size $L/2$. 
It shows strong oscillation but seems to saturate to a finite value.
(b) Time evolution of return probability.
The return probability remains at a finite value for long times.
Enhancement by the repulsions is observed.
This indicates that homogeneity produced by the repulsions enhances FMBL. 
The inset displays the high-resolution of temporal behavior of the return probability. We set $L=24$ system with open boundary condition.
}
\label{Fig3}
\end{figure}

\textit{Finite-density states.---}
The study of the two-particle systems revealed a very interesting phenomenon, that is,
$\Se$ develops substantially under time evolution, whereas the revival of the LIOMs takes place at the same time. 
We expect that this phenomenon comes from the fact that the energy eigenstates in the non-interacting
system are the same or quite similar even for $W^{+\dagger}_j|0\rangle$ and $W^{-\dagger}_k|0\rangle$,
and then, mixings of these states occur quite easily.
From this consideration, we expect that some specific localization behavior emerges in the system
at finite fillings. 
Properties of configurations may strongly depend on the filling factor.


\begin{figure}[t]
\begin{center} 
\includegraphics[width=7.5cm]{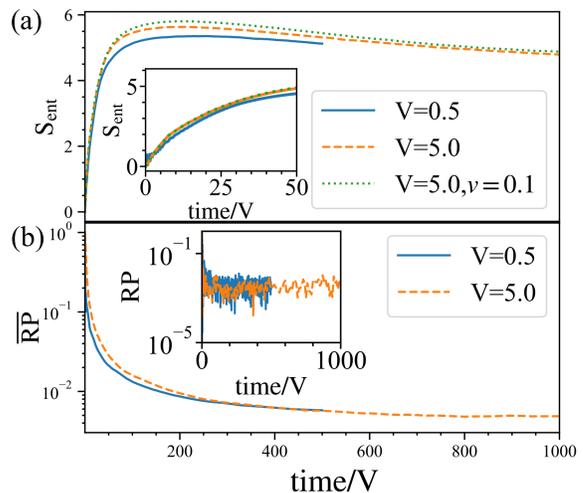} 
\end{center} 
\caption{
$1/4$-filling dynamics: (a) Time evolution of $S_{\rm ent}$ for subsystem with size $L/2$.
$\Se$ for extended state with finite inter-leg hopping $v=0.1$ is also shown.
 (b) Time evolution of return probability.
It decreases quite rapidly.
The inset displays high-resolution of temporal behavior of the return probability. We set $L=16$ system with open boundary condition.
}
\label{Fig4}
\end{figure}
With this expectation, we investigate many-particle dynamics by employing the TEBD method in TeNPy package \cite{TenPy}. 
The validity of the TEBD method is examined \cite{Supp} and sufficiently careful manipulations are given.
In what follows, we set open boundary condition. 
We first consider an initial state with particle filling $1/6$, $|I_{1/6}\rangle =\prod^{L/3-1}_{j=0}a^{\dagger}_{3j+1}|0\rangle$ and calculate 
$S_{\rm ent}$ for the subsystem of the left half of the ladder and 
return probability $|\langle \Psi(t)|I_{\alpha}\rangle|$.
In this particle filling, all particles initially start Aharanov-Bohm caging, and after that the particles interact
with each other. 
This behavior is similar to the two-particle systems studied in the above. 
The results obtained by TEBD are displayed in Fig.~\ref{Fig3}.
For early times ($\mbox {time}/V \lesssim 10^{1}$), $S_{\rm ent}$ increases and then saturates. 
The saturation value is independent of the strength of the repulsion, $V$, 
and comparable to the maximum of $S_{\rm ent}$ of 
the two-particle dynamics in Fig.~\ref{Fig2} (a), which is smaller than that of a conventional thermal (extended) 
many-body states. 
The return probability in Fig.~\ref{Fig3} (b) also shows an interesting behavior, that is, the PR clearly oscillates even for 
a long-time, and this character is also independent of the value of $V$. 
The memory of the initial state is conserved.
We expect that the origin of this phenomenon is the remnants of LIOM picture in the flat-band system, i.e., 
the $\ell$-bits survive in the presence of the interactions by the mechanism that we observed for 
the two-particle system. 
From the results of $S_{\rm ent}$ and return probability, the system is not thermalized and this indicates ergodicity breaking dynamics, 
expected by our small system-size calculations in \cite{KOI2020}. 
The above characters are analog of the dynamical behavior of the conventional MBL \cite{Huse}, 
but there is a difference, i.e., the $S_{\rm ent}$ does not exhibit a logarithmic increase.
We expect that this is a signature of the localization mechanism in the flat-band system, which 
works efficiently for the homogeneous states. 
In addition, it should be commented that in Rydberg atom, it has been observed that the revival dynamics occurs 
for a long time when the initial state is taken as Neel state \cite{Bernien}. 
There is a gauge-theoretical interpretation of this phenomenon \cite{Smith1,Smith2,PRD,Surace}. 
It is just a string inversion mechanism.
We observe similarity between the LIOMs and gauge-invariance constraint.
 
An initial state with particle filling $1/4$, $|I_{1/4}\rangle =\prod^{L/4-1}_{j=0}a^{\dagger}_{2j+1}|0\rangle$ 
is also studied.
The particles in this case interact with each other more strongly than the case of the $1/6$ filling (See \cite{Supp}).
The time evolution of this case, shown in Fig.~\ref{Fig4}, seems quite different from that of  the $1/6$-filling case
in Fig.~\ref{Fig3}.
It indicates that the $\ell$-bit picture is broken as a result of substantial modification of states. 
The $1/4$ filling is an extended state. 
Actually, the value of $S_{\rm ent}$ of the $1/4$-filling flat band with interactions saturates to
the same value of the band bending case constructed by inclusion of
a vertical hopping, $H_v=v\sum^{L}_{j=1}(a^\dagger_j b_j+\mbox{h.c.})$.

{\it Conclusion.---}
We clarified dynamical aspect, especially entanglement property, of the FMBL in the Creutz ladder model 
with interactions.
The LIOM obtained by $\ell$-bit (compact localized state) plays a major role in the appearance of the FMBL.
Starting with studying the fate of the Aharanov-Bohm caging with interactions for the two-particle dynamics, 
we investigated many-body dynamics by using the TEBD. 
The presence of the FMBL is determined by the balance of particle filling and the range of inter-particle interactions.
For a suitable choice of particle filling, 
the many-body dynamics with interactions exhibits non-thermalized behavior with low-entanglement growth and finite return probability. 
The $\ell$-bit picture and LIOMs in the flat-band system survive there.\\

This work was supported by the Grant-in-Aid for JSPS Fellows (No.17J00486).

\clearpage

\section*{Supplemental Material}

\subsection{One-particle system and its dynamics}

\begin{figure}[h]
\begin{center} 
\includegraphics[width=8cm]{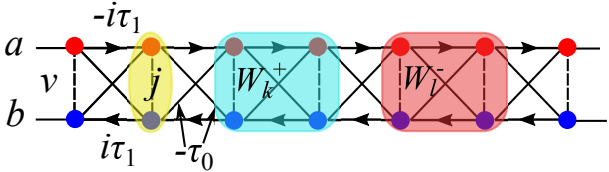} 
\end{center}
\caption{Schematic figure of the Creutz ladder model
}
\label{FigS1}
\end{figure}

We study dynamical behaviors of single-particle states for $H_{\rm flat}$ in (2) in the main text. 
The schematic figure of the model is given in Fig.~\ref{FigS1}.
The dynamics is obtained by the exact diagonalization \cite{Quspin}. 
Here, besides $H_{\rm flat}$ in (2) in the main text, 
we also add inter-leg hopping, defined by $H_v=v\sum_j(a^\dagger_j b_j+\mbox{h.c.})$. 
This hopping makes the band dispersive and destroys the localization properties. 
Here, we investigate the effects of $H_v$.
For the genuine system with Eq.~(2) in the main text, 
$K^{\pm}_j \equiv (W^{\pm}_j)^\dagger W^{\pm}_j$ are
perfect constants of motion, whereas $H_{\rm v}$ and $\tau_0\neq \tau_1$ break the conservation of the LIOMs.

\begin{figure}[t]
\begin{center} 
\includegraphics[width=8cm]{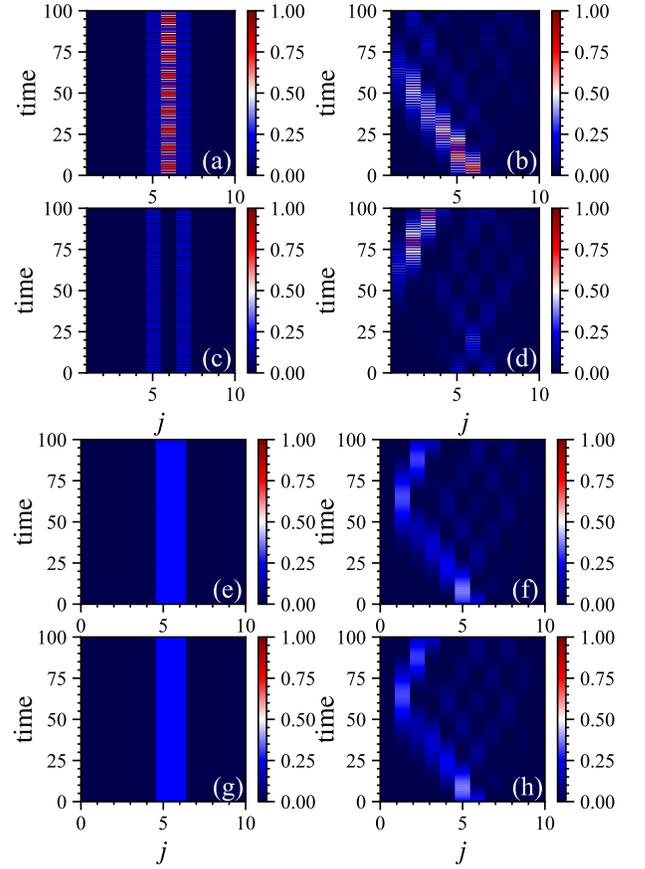} 
\end{center}
\caption{(a) Time evolution of $a$-particle density with the initial state $a^\dagger_{L/2}|0\rangle$
for the flat band. 
(b) Time evolution of $a$-particle for $v=0.1$. 
(c) Time evolution of $b$-particle density for the flat band.
(d) Time evolution of $b$-particle density for $v=0.1$.
(e) Time evolution of $\langle K^+_j\rangle$ for the flat band.
(f) Time evolution of $\langle K^+_j\rangle$ for $v=0.1$
(g) Time evolution of $\langle K^-_j\rangle$ for the flat band.
(h) Time evolution of $\langle K^-_j\rangle$ for $v=0.1$.
}
\label{FigS2}
\end{figure}

As the initial state, the single-particle state such as $a^\dagger_{L/2}|0\rangle$ is employed.
We study the time evolution of the particle densities and LIOMs, $\langle K^\pm_j\rangle$.
For the system $H_{\rm flat}$ in Eq.~(2),  
the a/b-particle densities during time evolution are quite stable as shown in Figs.~\ref{FigS2} (a) and (c), 
and the LIOM $\langle K^\pm_j\rangle$ remains constant as shown in Figs.~\ref{FigS2} (e) and (g). 
These results are just Aharanov-Bohm caging \cite{vidal0,vidal1}. 
On the other hand, once one switches on $H_{\rm v}$ or sets $\tau_1\neq \tau_0$, it breaks the flat-band structure, 
and induces instability of the LIOMs. 
As shown in Figs.~\ref{FigS2} (b), (d), and (f), 
the a- and b- particle densities spread over the whole system, and the LIOMs are not conserved.
However, it is interesting and instructive to see that region with the significant 
expectation values of the LIOMs, $\langle K^\pm_j\rangle$, moves without breaking its shape and bounces 
at the boundary.

\subsection{Analytical solution of dynamical entanglement entropy for small $V$}

In Fig.1 in the main text, we consider the two-particle dynamics with the initial state;
${1\over2}(a_4^\dagger+ib_4^\dagger)(a_6^\dagger+ib_6^\dagger)|0\rangle
={1\over2}(W^{+\dagger}_4-W^{-\dagger}_4)(W^{+\dagger}_6-W^{-\dagger}_6)|0\rangle$. 
Analytical solution of the dynamical entanglement entropy ($S_{\rm ent}$) for small $V$ can be obtained 
by applying the simple perturbation theory for degenerate states. 
Here, we consider the flat-band case.

The initial state is a superposition of the two-particle states, 
$W^{+\dagger}_4W^{+\dagger}_{6}|0\rangle$, $W^{+\dagger}_4W^{-\dagger}_{6}|0\rangle$,
$W^{-\dagger}_4W^{+\dagger}_{6}|0\rangle$ and 
$W^{-\dagger}_4W^{-\dagger}_{6}|0\rangle$.
For $H_{\rm flat}$, the states $W^{+\dagger}_4W^{-\dagger}_{6}|0\rangle$ and 
$W^{-\dagger}_4W^{+\dagger}_{6}|0\rangle$ are degenerate with the vanishing energy. 
Thus, to obtain the dynamics 
$S_{\rm ent}$, we need to calculate the two-particle wave function at time $t$. 
To this end, we calculate perturbated energies and the corresponding perturbated eigenstates 
for the two-particle states. 
For the Hamiltonian $H_{\rm flat}+ H_{\rm I}$ ($H_{\rm flat}$ is non-perturbative and $H_{\rm I}$ is perturbative), 
the first order perturbation theory gives the following two-particle approximated non-degenerate eigenstates
straightfowardly,
\be
|1\rangle &=& W^{+\dagger}_4W^{+\dagger}_{6}|0\rangle \:\:\mbox{with}\:\: E_1=4\tau_{0},\nonumber\\
|2\rangle &=& {1\over\sqrt{2}}(W^{+\dagger}_4W^{+\dagger}_{6}|0\rangle +W^{+\dagger}_4W^{+\dagger}_{6}|0\rangle) \:\:\mbox{with}\:\: E_2=0,\nonumber\\
|3\rangle &=& {1\over\sqrt{2}}(W^{+\dagger}_4W^{+\dagger}_{6}|0\rangle- W^{+\dagger}_4W^{+\dagger}_{6}|0\rangle) \:\:\mbox{with}\:\: E_3=V/4,\nonumber\\
|4\rangle &=& W^{+\dagger}_4W^{+\dagger}_{6}|0\rangle \:\:\mbox{with}\:\: E_4=-4\tau_{0}.
\label{eigenstates}
\ee
Here, $|\alpha\rangle$ ($\alpha=1,2,3,4$) are perturbated eigenstates from the states, 
$W^{+\dagger}_4W^{+\dagger}_{6}|0\rangle$, $W^{+\dagger}_4W^{-\dagger}_{6}|0\rangle$,
$W^{-\dagger}_4W^{+\dagger}_{6}|0\rangle$ and 
$W^{-\dagger}_4W^{-\dagger}_{6}|0\rangle$.

By using the above eigenstates, we get the time-evolved state as follows, 
\be
|\psi(t)\rangle&=&e^{-i(H_{\rm flat}+ H_{\rm I})t}|{1\over2}(a_4^\dagger+ib_4^\dagger)(a_6^\dagger+ib_6^\dagger)|0\rangle\nonumber\\
&\sim& {1\over2}e^{-i(4\tau +{V\over8})t}|W^+_4W^+_6\rangle-{1\over2}|W^+_4W^-_6\rangle\nonumber \\
&&-{1\over2}|W^-_4W^+_6\rangle+{1\over2}e^{i(4\tau -{V\over8})t}|W^-_4W^-_6\rangle.
\label{psi_t}
\ee
From this time-evolved wave function, the partial density matrix $\rho_{L}$ is obtained as follows
by tracing out the right half of the system,
\be
\rho_{L}={1 \over 2}\begin{pmatrix} 1& e^{-4i\tau t}\rm{cos}({V\over8}) \\ 
e^{4i\tau t}\rm{cos}({V\over8}) & 1 \end{pmatrix}
\label{sup_rho}
\ee
Therefore, the time evolution of $S_{\rm ent}$ defined by $S_{\rm ent}=-{\rm Tr}[\rho_{L}\log \rho_{L}]$ is given by 
\be
S_{\rm ent}(t) &=&-{1+|\cos{V\over8}t|\over2}\rm{log}\biggl[{1+|\cos{V\over8}t|\over2}\biggr]\nonumber\\
&&-{1-|\cos{V\over8}t|\over2}\rm{log}\biggl[{1-|\cos{V\over8}t|\over2}\biggr].
\label{analytical_EE}
\ee
This solution exhibits an oscillation with the period $8\pi/V$, which is plotted in Fig. 1 (a) in the main text. 

\subsection{Validity and technical comments on numerical TEBD simulations}

For many-body dynamics, we employ the time-evolving block decimation (TEBD) method 
by using TeNPy package \cite{TenPy}. 
By comparing the method to the exact diagonalization of Quspin package \cite{Quspin}, 
we verify the validity of the TEBD method for the flat-band system in the main text.

\begin{figure}[t]
\begin{center} 
\vspace{1cm}
\includegraphics[width=8cm]{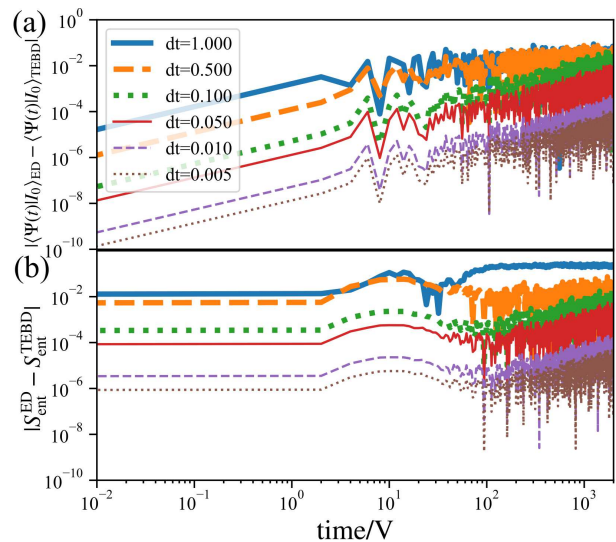} 
\end{center} 
\caption{Comparison of TEBD and exact diagonalization calculations: (a) Time evolution of the difference of the return probability, defined by $|\langle \Psi(t)|I_{0}\rangle_{\rm ED}-\langle \Psi(t)|I_{0}\rangle_{\rm TEBD}|$. 
(b) Time evolution of the difference of $S_{\rm ent}$, defined by $|S^{\rm ED}_{\rm ent}-S^{\rm TEBD}_{\rm ent}|$.}
\label{FigS3}
\end{figure}

We consider the system $H_{\rm flat}+ H_{\rm I}$ with $\tau_{0}=1$ and $V=2.0$.
The system size is $L=10$ (20 sites) and employ the open boundary condition. 
We set an initial state, $|I_0\rangle = a_1^\dagger a_3^\dagger a_5^\dagger a_7^\dagger a_9^\dagger|0\rangle$, 
and measure the return probability and the entanglement entropy, $S_{\rm ent}$, for 
the subsystem of the left half of the ladder as defined in the main text. 
Here, as a technical aspect, it should be noted that for the TEBD calculation, 
we add very small band bending by including the cross hopping term with $\tau_{0}=10^{-8}\tau_{1}$ 
to avoid the zero-energy effects in exact flat band situation, 
but such a small modification has little effects on the dynamics of the physical quantities of interest. 
We show this fact by comparing with the exact diagonalization, which does not include such a band bending.

Except for the modification in the TEBD method, we calculate the dynamics by the TEBD and exact diagonalization 
in the same condition. Figure \ref{FigS3} (a) and (b) are the comparison between both calculations 
for various intervals of time step, $dt$. For smaller $dt$, Both differences for the return probability and $S_{\rm ent}$ decrease. 
For the difference of the return probability, the value is almost below $\mathcal{O}(10^{-4})$ for $dt=0.01$ 
within our target time interval $~10^{3}$. The error does not affect our focusing physics. 
For the cauclulations of $S_{\rm ent}$, the same order of accuracy is verified
as shown in Fig.~\ref{FigS3} (b). The value of the difference is almost below 
$\mathcal{O}(10^{-4})$ for $dt=0.01$ within our target time interval $10^{3}$. 
Therefore, the error also does not affect our focusing physics. 
From the above studies on the accuracy of the numerical methods, for
 our TEBD calculations in the main text, we set $dt=0.01$ and included small band bending by $\tau_{0}=10^{-8}\tau_{1}$.


\begin{thebibliography}{99}
\bibitem{remark1}
In observing dynamics of the system the fourth-order moment of LIOMs is efficient to estimate the deviation from the LIOM picture of the initial state.
\bibitem{remark2}
$j$-subsystem is the subsystem of the left half of the ladder and $k$-subsystem is of the right.
\end{thebibliography}
\end{document}